\documentclass[aps,showpacs,superscriptaddress,preprint,nofootinbib]{revtex4-2}
\usepackage{latexsym,graphicx,epsfig,psfrag}

\usepackage{epstopdf}
\usepackage{amssymb,amsmath,amsxtra,amsfonts}
\usepackage{bm}

\usepackage{longtable}
\usepackage{multirow}
\usepackage{booktabs}
\usepackage{array}
\usepackage{wrapfig}
\usepackage{color}
\usepackage{titlesec}

\titleformat{\section}{\large\bfseries}{\thesection}{1em}{}

\newcommand{\bea}{\begin{eqnarray}}
\newcommand{\ena}{\end{eqnarray}}
\newcommand{\be}{\begin{equation}}
\newcommand{\en}{\end{equation}}
\newcommand{\nn}{\nonumber\\}
\newcommand{\ed}{\end{document}} 
\newcommand{\Tr}{\mbox{\rm{tr}}}

\begin{document}

\title{Probing the ATOMKI X17 vector boson using Dalitz decays $\bm{V\to Pe^+e^-$}}

\homepage{Supported by Ho Chi Minh City University of Technology and Education under Grant T2024-65}

\author{C.~T.~Tran}
\email{E-mail: thangtc@hcmute.edu.vn}
\affiliation{
	\hbox{Department of Physics, HCMC University of Technology and Education, }\\
	Vo Van Ngan 1, 700000 Ho Chi Minh City, Vietnam}

\author{M.~A.~Ivanov}
\affiliation{Bogoliubov Laboratory of Theoretical Physics, 
Joint Institute for Nuclear Research, 141980 Dubna, Russia}

\author{A.~T.~T.~Nguyen}
\affiliation{
	Department of Physics, HCMC University of Education, An Duong Vuong 280, 700000 Ho Chi Minh City, Vietnam}

\begin{abstract}
Recent anomalies observed in $e^+e^-$ nuclear transitions of $^8 \textrm{Be}$, $^4 \textrm{He}$, and $^{12} \textrm{C}$ by the ATOMKI collaboration may hint at the existence of a vector boson with a mass around 17~MeV, referred to as X17. If it exists, this boson would also affect similar processes in particle physics, including the Dalitz decays of vector mesons. Recently, the BESIII collaboration measured the Dalitz decay $D^{*0}\to D^0e^+e^-$ for the first time and reported a $3.5\sigma$ excess over the theoretical prediction based on the vector meson dominance (VMD) model. This excess may be another signal of the X17. In this study, we investigate the possible effects of the X17 on the Dalitz decays $D^*_{(s)}\to D_{(s)}e^+e^-$, $B^*_{(s)}\to B_{(s)}e^+e^-$, and $J/\psi\to \eta_ce^+e^-$. The required hadronic form factors are calculated within the framework of our covariant confined quark model, without relying on heavy quark effective theory or the VMD model. We present predictions for the Dalitz decay widths and the ratios $R_{ee}(V)\equiv \Gamma(V\to Pe^+e^-)/\Gamma(V\to P\gamma)$ within the Standard Model and in several new physics scenarios involving modifications due to the X17. Our results are compared with other theoretical calculations. 
\end{abstract}

\pacs{13.25.Ft, 13.20.-v, 12.39.Ki}
\keywords{covariant confined quark model, Dalitz decay, X17}

\maketitle
\newpage

\section{Introduction}
\label{sec:intro}
In 2015, the ATOMKI collaboration reported an anomaly in the M1 transition of the 18.15~MeV isoscalar excited state ($J^\pi = 1^+, T = 0$) to the ground state ($J^\pi = 0^+, T = 0$) in $^8\textrm{Be}$~\cite{Krasznahorkay:2015iga, Krasznahorkay:2019lgi}. The excited state of $^8\textrm{Be}$ was created by firing proton beam at thin strip foils of $^7\textrm{Li}$. The first then decayed to the ground state and produced pairs of electrons and positrons. The $e^+e^-$ emission is referred to as the internal pair creation (IPC). The angular correlation of the IPC process (IPCC) was measured at various opening angles $\theta$ in the laboratory rest frame. Quantum electrodynamics (QED) predicts the IPCC to drop rapidly and monotonically with the increase in the opening angle~\cite{Rose:1949zz, Schluter:1981cjo}. However, a deviation was observed at the proton beam energy $E_p = 1.10~\textrm{MeV}$ and at the angle $\theta \approx 140^\circ$ with a $6.8\sigma$ significance. The deviation appeared as a ``bump" in the angular correlation spectrum, suggesting an additional contribution to the $e^+e^-$ production from a hypothetical bosonic particle of mass $\approx 17~\textrm{MeV}$, namely X17, apart from the main photon contribution. 

After the first observation of the anomaly in the $^7\textrm{Li}(\textrm{p},e^+e^-)^8\textrm{Be}$ nuclear reaction~\cite{Krasznahorkay:2015iga}, the ATOMKI collaboration further studied the reactions $^3\textrm{H}(\textrm{p},e^+e^-)^4\textrm{He}$~\cite{Krasznahorkay:2021joi} and $^{11}\textrm{B}(\textrm{p},e^+e^-)^{12}\textrm{C}$~\cite{Krasznahorkay:2022pxs} and reported the observation of similar anomalies. The $e^+e^-$ excesses in the decays of excited $^4\textrm{He}$ and $^{12}\textrm{C}$ were observed at different opening angles compared to the case of excited $^8\textrm{Be}$, but the inferred mass of the unknown particle remained consistent with the $^8\textrm{Be}$ result. 
\begin{table}[htbp]
	\caption{ATOMKI anomaly summary.}\label{tab:Atomki-sum}
	\renewcommand{\arraystretch}{0.7}
	\begin{ruledtabular}
		\begin{tabular}{cccccc}
			Transition & Opening angle $\theta$ & $E_p$ (MeV) & Best fit  $m_X$ (MeV) & Confidence & Ref. \\
			\hline
			$^7\textrm{Li}(\textrm{p},e^+e^-)^8\textrm{Be}$ &	$\approx 140^\circ$ & 1.10 & $16.70\pm 0.35 (\textrm{stat})\pm 0.50 (\textrm{sys})$ & $6.8\sigma$ & \cite{Krasznahorkay:2015iga} \\[1.5ex]
			$^3\textrm{H}(\textrm{p},e^+e^-)^4\textrm{He}$ & $\approx 115^\circ$ & 0.51 & $17.01 \pm 0.12 (\textrm{stat})\pm 0.21 (\textrm{sys}) $ & $7.3\sigma$ &   \cite{Krasznahorkay:2021joi}\\
			& & 0.61 & $16.88 \pm 0.16 (\textrm{stat})\pm 0.21 (\textrm{sys}) $ & $6.6\sigma$ &\\
				& & 0.90 & $16.68 \pm 0.30 (\textrm{stat})\pm 0.21 (\textrm{sys}) $ & $8.9\sigma$ &\\[1.5ex]
			$^{11}\textrm{B}(\textrm{p},e^+e^-)^{12}\textrm{C}$ & $\approx 155^\circ - 160^\circ$ & 1.50 & $16.81 \pm 0.15 (\textrm{stat})\pm 0.20 (\textrm{sys}) $ & $3.0\sigma$ & \cite{Krasznahorkay:2022pxs}\\
				& & 1.70 & $16.93 \pm 0.08 (\textrm{stat})\pm 0.20 (\textrm{sys}) $ & $7.0\sigma$ &\\
			& & 1.88 & $17.13 \pm 0.10 (\textrm{stat})\pm 0.20 (\textrm{sys}) $ & $8.0\sigma$ &\\
			& & 2.10 & $17.06 \pm 0.10 (\textrm{stat})\pm 0.20 (\textrm{sys}) $ & $3.0\sigma$ &\\
		\end{tabular}
	\end{ruledtabular}
\end{table}     
A brief summary of the ATOMKI's results is given in Table~\ref{tab:Atomki-sum}. Independent experimental attempts from other collaborations to test the ATOMKI anomaly have been done. Based on data samples with total statistics corresponding to $8.4\times 10^{10}$ electrons collected in 2017 and 2018, the NA64 collaboration at CERN searched for the X17 produced in the bremsstrahlung reaction $e^-Z\to e^-ZX$. No evidence was found and an improved limit on the $X-e^-$ coupling was set to be $1.2\times 10^{-4}\lesssim \varepsilon_e \lesssim 6.8\times 10^{-4}$~\cite{NA64:2018lsq,NA64:2019auh, NA64:2020xxh}. In 2023, the MEG~II collaboration at PSI reported no significant signal above the expected background in the $e^+e^-$ angular correlation spectrum that would indicate the presence of the X17 boson~\cite{MEGII:2024urz}. However, their results were found to be compatible with the ATOMKI's within $1.5\sigma$. Meanwhile, a group at VNU University of Science reported the observation of the anomaly in the $^7\textrm{Li}(\textrm{p},e^+e^-)^8\textrm{Be}$ nuclear reaction at proton beam energy 1.225~MeV and opening angle around $135^\circ$. The mass was found to be $m_X = 16.66\pm 0.47(\textrm{stat})\pm 0.35 (\textrm{sys})$~MeV with a confidence above $4\sigma$~\cite{Anh:2024req}.  Recently, the PADME collaboration has performed a dedicated search for the X17 boson via $e^+e^-$ annihilation and found an excess of events over the predicted background expectation with a local significance of $2.5\sigma$ for an X17 mass of 16.90~MeV~\cite{Bossi:2025ptv, PADME:2025dvz}. 

The ATOMKI anomaly has attracted a great deal of attention in the particle physics community and has led to a large number of theoretical studies searching for possible explanations. One promising solution to the anomaly is the ``protophobic" vector boson explanation proposed by Feng \textit{et al.}~\cite{Feng:2016jff, Feng:2016ysn} (see, however, Ref.~\cite{Zhang:2020ukq}). They also pointed out that the X17 boson could not be a scalar due to angular momentum conservation in the decay $^8\textrm{Be}^*\to\,  ^8\textrm{Be}+e^+e^-$.  Kozaczuk \textit{et al.} considered the possibility of the X17 being a light axial vector boson and found that such a scenario is consistent with experimental data~\cite{Kozaczuk:2016nma} (see also Ref.~\cite{Barducci:2022lqd}). Ellwanger and Moretti studied the pseudoscalar boson explanation in Ref.~\cite{Ellwanger:2016wfe}. In Ref.~\cite{Kirpichnikov:2020tcf}, Kirpichnikov, Lyubovitskij, and Zhevlakov performed a global analysis of possible impacts of hidden sub-GeV bosons with different spin-parity quantum numbers (including the X17 state) on the ATOMKI anomaly, $(g-2)_\mu$ anomaly, proton-charge-radius puzzle, and electric dipole moments of fermions. A large variety of models following the above mentioned directions has been proposed and studied~\cite{Pulice:2019xel, Seto:2020jal, Nomura:2020kcw, DiLuzio:2025ojt,Barducci:2025hpg}. A detailed review of the X17 anomaly can be found in Ref.~\cite{Alves:2023ree}.

If the X17 exists, it would affect similar processes in particle physics, including the Dalitz decays of vector mesons. Recently, the BESIII collaboration measured the Dalitz decay $D^{*0}\to D^0e^+e^-$ for the first time~\cite{BESIII:2021vyq} and reported a $3.5\sigma$ excess over the theoretical prediction based on the vector meson dominance (VMD) model~\cite{Sakurai:1960ju, Landsberg:1985gaz}. This excess may be another signal of the X17. The idea of using the Dalitz decays to probe for new physics beyond the Standard Model (SM) is not new. For instance, the Dalitz decays $J/\psi\to P\ell^+\ell^-$ have been proposed~\cite{Fu:2011yy} as a probe for dark photon searches at the BESIII experiment, where a large $J/\psi$ data sample has been accumulated~\cite{BESIII:2018aao}. In the light of the ATOMKI anomaly, Castro and Quintero proposed tests of the anomaly using the Dalitz decays $D^*_{(s)}\to D_{(s)}e^+e^-$ and $B^*_{(s)}\to B_{(s)}e^+e^-$~\cite{Castro:2021gdf}, assuming that X17 is a vector boson. Ban \textit{et al.}~\cite{Ban:2020uii} studied various strategies for searching the X17 boson in the Dalitz decay $J/\psi\to \eta_c\ell^+\ell^-$ at BESIII and Belle~II. Recently, Lee \textit{et al.}~\cite{Lee:2025lwv} performed a fit for the couplings between the X17 and quarks using experimental data from several Dalitz channels, including $\psi(2S)\to \eta_c\ell^+\ell^-$, $\phi\to \eta\ell^+\ell^-$, $D_s^*\to D_s\ell^+\ell^-$, and $D^{*0}\to D^0\ell^+\ell^-$. Note that in Refs.~\cite{Castro:2021gdf, Ban:2020uii, Lee:2025lwv} the authors relied on the VMD model and the heavy quark effective theory (HQET) to calculate the hadronic form factors. Also, there were some disagreements on the prediction for $D_s^*\to D_s\ell^+\ell^-$ between Ref.~\cite{Castro:2021gdf} and Ref.~\cite{Lee:2025lwv}.

In this study, we aim at providing independent predictions for the Dalitz decays $D^*_{(s)}\to D_{(s)}e^+e^-$, $B^*_{(s)}\to B_{(s)}e^+e^-$, and $J/\psi\to \eta_ce^+e^-$ in the SM, as well as in the presence of the X17. We follow the authors of Refs.~\cite{Castro:2021gdf, Lee:2025lwv} and assume that X17 is a vector boson. We use several favored combinations of the coupling parameters between the X17 and quarks to calculate the X17's effects on the widths of these decays and compare our results with those given in Refs.~\cite{Castro:2021gdf, Lee:2025lwv}. The hadronic form factors required for the calculation of the Dalitz decays are obtained in the Covariant Confined Quark Model (CCQM) which has been developed by our group. These form factors are calculated directly in our model for the entire range of momentum transfer without any extrapolations, heavy-quark limit or VMD assumptions. Therefore, our results would shed more light on the understanding of the X17 effects and help probe for the X17 boson in the Dalitz decays of vector mesons.

The rest of the paper is organized as follows. In Section~\ref{sec:model} we briefly introduce the Covariant Confined Quark Model as a tool for hadronic calculation. Section~\ref{sec:formalism} is devoted for the description of the  Dalitz decays in terms of the Feynman diagrams, invariant matrix elements, and form factors. Here we consider both the contributions from the photon (within the SM) and from the X17 vector boson (beyond the SM). Numerical results are presented in Section~\ref{sec:result}. A short description of the form-factor evaluation within the CCQM is also provided. Finally, a brief summary is given in Section~\ref{sec:sum}.
\section{Model}
\label{sec:model}
The covariant confined quark model is a framework based on quantum field theory that describes hadronic bound states as quantum fields interacting with their quark constituents. This interaction is defined by a Lagrangian. For a meson $M$, the Lagrangian is written as
\begin{equation}
	\mathcal{L}_{\mathrm{int}}(x) =g_M M(x)J(x)+\mathrm{H.c.},\qquad
	J(x) = \int dx_1\int dx_2 F_M(x;x_1,x_2)[\bar{q}_2(x_2)\Gamma_M q_1(x_1)],
\end{equation}  
where $g_M$ is the meson-quark coupling constant, $\Gamma_M$ is the Dirac matrix with quantum numbers specific to the meson $M$, and $F_M(x;x_1,x_2)$ is the vertex function that effectively describes the meson's size and the relative positioning of the quarks and the meson as a whole. The vertex function has the form
\begin{equation}
	F_M(x;x_1,x_2) = \delta^{(4)}(x-\omega_1 x_1-\omega_2 x_2)\Phi_M[(x_1-x_2)^2],
\end{equation} 
where $\omega_{i}=m_{q_i}/(m_{q_1}+m_{q_2})$ is the mass fraction of the quarks. The function $\Phi_M[(x_1-x_2)^2]$ is assumed to be Gaussian for simplicity, and takes the following form in momentum space
\begin{equation}
	\widetilde{\Phi}_M(-p^2)=\exp(p^2/\Lambda^2_M).
	\label{eq:vertexf}
\end{equation}
Here, $\Lambda_M$ is a free parameter of the model, referred to as the size parameter of the meson $M$. It is important to note that the specific mathematical form chosen for $\widetilde{\Phi}_M(-p^2)$ is not critical, as long as it decreases rapidly enough at high momentum values (in the Euclidean space) to ensure the ultraviolet finite of Feynman diagrams.

The normalization of particle-quark vertices is provided by the compositeness 
condition~\cite{Salam:1962ap,Weinberg:1962hj}
\begin{equation}
	Z_M = 1 - \Pi^\prime_M(m^2_M) = 0,
	\label{eq:compositeness}
\end{equation}
where $Z_M$ is the meson's wave function renormalization constant and $\Pi'_M$ is the derivative of the meson's mass function.
\begin{figure}[htbp]
	\includegraphics[width=0.40\textwidth]{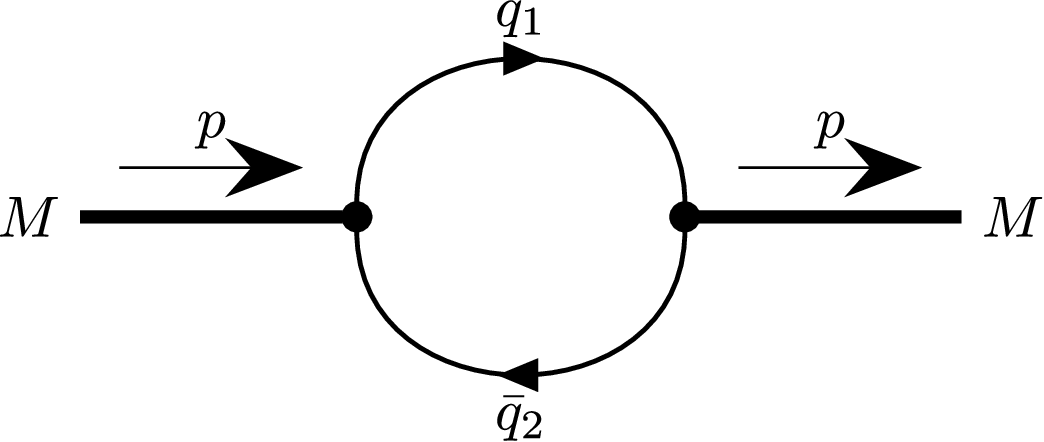}
	\caption{One-loop self-energy diagram for a meson.}
	\label{fig:mass}
\end{figure}
The meson mass function $\Pi_M(p)$ is calculated using a one-loop Feynman diagram representing the meson's self-energy (see Fig.~\ref{fig:mass}). The specific forms for pseudoscalar $(\Pi_P)$ and vector $(\Pi_V)$ mesons read
\begin{eqnarray}
	\Pi_P(p) &=& 3g_P^2 \int\!\! \frac{dk}{(2\pi)^4i}\,\widetilde\Phi^2_P \left(-k^2\right)
	\Tr\left[ S_1(k+w_1p)\gamma^5 S_2(k-w_2p)\gamma^5 \right],\\
	\label{eq:Pmass-Pseudoscalar}
	\Pi_V(p) &=& g_V^2 \left[g^{\mu\nu} - \frac{p^{\mu}p^{\nu}}{p^2}\right] 
	\int\!\! \frac{dk}{(2\pi)^4i}\,\widetilde\Phi^2_V \left(-k^2\right)
	\Tr\left[ S_1(k+w_1p)\gamma_{\mu} S_2(k-w_2p)\gamma_{\nu} \right],
\end{eqnarray}
where the free quark propagator has been used
\begin{equation}
	S_i(k) = \frac{1}{m_{q_i} - \not\! k - i\epsilon}= \frac{m_{q_i}+\not\! k}{m^2_{q_i} - k^2 - i\epsilon}.
	\label{eq:prop}
\end{equation}

The CCQM has several adjustable parameters: the constituent quark masses $m_q$, the size parameters for hadrons $\Lambda_H$, and a universal cutoff parameter $\lambda$ that accounts for the confinement of quarks within hadrons at low energies. These parameters are determined by fitting the model's predictions to experimental data. Once these parameters are fixed, the CCQM can be used to perform hadronic calculations across the entire physical range of momentum transfer without needing to extrapolate. A key advantage of the model is its ability to consistently describe not only mesons~\cite{Faessler:2002ut, Ivanov:2006ni, Ivanov:2020iad, Ivanov:2019nqd, Tran:2024phq, Tran:2025ulp} but also baryons~\cite{Gutsche:2013pp, Gutsche:2018nks, Groote:2021ayy}, tetraquarks~\cite{Dubnicka:2011mm, Goerke:2016hxf}, and other multiquark states~\cite{Gutsche:2017twh, Soni:2020sgn}. The values of the model parameters used in this work are provided in Tables~\ref{tab:size_parameter} and~\ref{tab:quark_mass}. Other physical quantities and model-independent parameters are taken from the Particle Data Group~\cite{ParticleDataGroup:2024cfk}. 

\begin{table}[ht]
	\caption{Meson size parameters (in GeV).}\label{tab:size_parameter}
	\renewcommand{\arraystretch}{0.7}
	\begin{ruledtabular}
		\begin{tabular}{cccccccccc}
			$\Lambda_{D}$ & $\Lambda_{D_s}$ & $\Lambda_B$ & $\Lambda_{B_s}$ & $\Lambda_{\eta_c}$ & $\Lambda_{D^*}$ & $\Lambda_{D_s^*}$ & $\Lambda_{B^*}$ & $\Lambda_{B_s^*}$ & $\Lambda_{J/\psi}$ \\
			\hline
			1.600 & 1.748 & 1.963 & 2.050 & 3.777 & 1.529 & 1.558 & 1.805 & 1.794 & 1.738 
		\end{tabular}
	\end{ruledtabular}
\end{table}
\begin{table}[ht]
	\caption{Quark masses and infrared cutoff parameter (in GeV).}\label{tab:quark_mass}
	\renewcommand{\arraystretch}{0.7}
	\begin{ruledtabular}
		\begin{tabular}{ccccc}
			$m_{u/d}$ & $m_s$ &  $m_c$ &  $m_b$ & $\lambda$ \\
			\hline
			0.241 & 0.428 & 1.672 & 5.046 & 0.181
		\end{tabular}
	\end{ruledtabular}
\end{table}

To estimate theoretical uncertainties within the CCQM, we note that we employed the MINUIT program, which finds the best parameter values by minimizing the $\chi^2$
function when fitting the model to data. Instead of a complex error propagation analysis, we adopt a simpler, though less precise, method for estimating errors in our physical predictions. We proceeded with calculations using only these best-fit parameter values. Upon comparison, we noted that the model's fitted values typically deviated from experimental data by about 5\%$-$10\%. Since our calculations of hadronic quantities are analogous to those for leptonic and electromagnetic decay constants used in the fit, we estimate the error in the CCQM predictions to be approximately 10\%.  

\section{Formalism}
\label{sec:formalism}
Within the SM, the Dalitz decay has been calculated previously in many theoretical studies (see, e.g., Refs.~\cite{Branz:2009cd, Gu:2019qwo, Tan:2021clg}). We therefore do not repeat the calculation and just write down the well known expression for the decay width distribution as
\begin{eqnarray}
	\frac{d\Gamma(V\to P
		\ell^+\ell^-)}{dq^2} &=& \frac{\alpha_{\textrm{em}}}{3\pi q^2}\sqrt{1-\frac{4m^2_\ell}{q^2}}\left(1+\frac{2m^2_\ell}{q^2}\right)\left[1-\frac{q^2}{(m_V-m_P)^2}\right]^{\frac32}\left[1-\frac{q^2}{(m_V+m_P)^2}\right]^{\frac32}\nonumber\\
	&\times& \Gamma(V\to P\gamma)	|F^\gamma_{VP}(q^2)|^2,\nonumber\\
	&\equiv& [\textrm{QED}(q^2)]\times |F^\gamma_{VP}(q^2)|^2.
	\label{eq:dGamSM}
\end{eqnarray}
The physical range for the momentum transfer is given by $4m^2_\ell \leq q^2 \leq (m_V-m_P)^2$.
\begin{figure}[htbp]
	\begin{tabular}{lr}
		\includegraphics[width=0.5\textwidth]{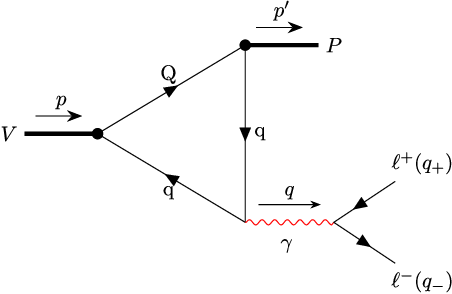}
		&
		\includegraphics[width=0.5\textwidth]{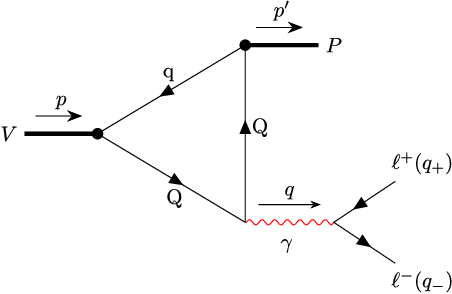}
	\end{tabular}
	\caption{Feynman diagrams for Dalitz decays $V\to P\ell^+\ell^-$.}
	\label{fig:Dalitz}
\end{figure}

In the CCQM, the transition form factor $F^\gamma_{VP}(q^2)$ is calculated based on the Feynman diagrams in Fig.~\ref{fig:Dalitz}, which describe direct photon emission from the constituent quarks. Note that within the CCQM, resonance diagrams corresponding to photon emission via intermediate $V^\prime\to\gamma^*$ transitions are also allowed (see, e.g., Ref.~\cite{Branz:2009cd}). However, since the mass splittings $\Delta m(VP)\equiv m_V-m_P$ for the decays studied in this paper are too small ($\Delta m(D^*D)\approx 140~\textrm{MeV}$, $\Delta m(B^*B)\approx 50~\textrm{MeV}$, and $\Delta m(J/\psi \eta_c)\approx 110~\textrm{MeV}$) compared to masses of the vector resonances $\rho(770)$, $\omega(782)$, $\phi(1020)$, and $J/\psi(3096)$, the contributions from resonance diagrams are negligibly small, and are well below the estimated inherent error of the model predictions ($\sim 10\%$). We therefore only consider the triangle diagrams in this study.  
Diagrams describing the contribution of the X17 boson to the Dalitz decays are obtained from those in Fig.~\ref{fig:Dalitz} by replacing the photon by the X17 boson.

The quark-photon coupling is described by the interaction Lagrangian
\begin{equation}
	\mathcal{L}^{\mathrm{int}}_{\mathrm{em}}(x) = eA_\mu(x)J_{\mathrm{em}}^\mu(x),\qquad J_{\mathrm{em}}^\mu(x) = e_Q\bar{Q}(x)\gamma^\mu Q(x)
	+ e_q\bar{q}(x)\gamma^\mu q(x),
\end{equation}
where $e_Q$ and $e_q$ are the charges of heavy and light quarks in units of $e$. Similarly, for the interaction between the quarks and the X17 boson one has
\begin{equation}
	\mathcal{L}^{\mathrm{int}}_{X}(x) = eX_\mu(x)J_{X}^\mu(x),\qquad J_{X}^\mu(x) = \varepsilon_Q\bar{Q}(x)\gamma^\mu Q(x)
	+ \varepsilon_q\bar{q}(x)\gamma^\mu q(x),
\end{equation}
where $\varepsilon_{Q,q}$ are the coupling constants between the X17 boson and quarks of various flavors.

The invariant matrix elements of the Dalitz decay $V(p,\epsilon_V)\to P(p^\prime)e^+(q_+)e^-(q_-)$ are given by~\cite{Lee:2025lwv}
\begin{eqnarray}
	\label{eq:M}
	i\mathcal{M}^\gamma &=& T_\mu^\gamma \frac{-ig^{\mu\nu}}{q^2+i\epsilon} (-ie)\bar{u}(q_-)\gamma_\nu v(q_+),\nn
	i\mathcal{M}^X &=& T_\mu^X \frac{-i(g^{\mu\nu}-q^\mu q^\nu/m_X^2)}{q^2-m_X^2+i m_X\Gamma_X} (-ie\varepsilon_e)\bar{u}(q_-)\gamma_\nu v(q_+),
\end{eqnarray}
where $e\varepsilon_e$ is the coupling constant between the X17 boson and electron (positron), and $T_\mu^{\gamma,X}$ are the hadronic amplitudes of the transitions $V\to P\gamma^*$ and $V\to PX^*$, respectively. The hadronic transition amplitudes can be parametrized in terms of the transition form factors $g_{VP\gamma}(q^2)$ and $g_{VPX}(q^2)$ as follows:
\begin{eqnarray}
	\label{eq:T}
	T_\mu^\gamma &\equiv& \left\langle P(p^\prime)|J_\mu^{\textrm{em}}|V(p,\epsilon_V) \right\rangle = ieg_{VP\gamma}(q^2) \epsilon_{\mu\nu\alpha\beta}\epsilon_V^\nu p^\alpha p^{\prime\beta},\nn
	T_\mu^X &\equiv& \left\langle P(p^\prime)|J_\mu^{X}|V(p,\epsilon_V) \right\rangle = ieg_{VPX}(q^2) \epsilon_{\mu\nu\alpha\beta}\epsilon_V^\nu p^\alpha p^{\prime\beta}.
\end{eqnarray}

The differential decay rates induced by photon and X17 boson are given by
\begin{eqnarray}
	\label{eq:dGamG}
	\frac{d\Gamma^\gamma}{dq^2} &=& \frac{\alpha^2_{\textrm{em}}}{72\pi m_V^3} g^2_{VP\gamma}(q^2)\frac{1}{q^2}\left(1+\frac{2m_e^2}{q^2}\right)\sqrt{1-\frac{4m_e^2}{q^2}}\lambda^{3/2}(m_V^2,m_P^2,q^2),\\
	\frac{d\Gamma^X}{dq^2} &=& \frac{\alpha^2_{\textrm{em}}\varepsilon_e^2}{72\pi m_V^3} g^2_{VPX}(q^2)\frac{q^2}{(q^2-m_X^2)+m_X^2\Gamma_X^2}\left(1+\frac{2m_e^2}{q^2}\right)\sqrt{1-\frac{4m_e^2}{q^2}}\nn
	&\times&\lambda^{3/2}(m_V^2,m_P^2,q^2),
	\label{eq:dGamX}
\end{eqnarray}
where $\lambda(x,y,z) \equiv x^2+y^2+z^2-2(xy+yz+zx)$ is 
the K{\"a}ll{\'e}n function.  

By integrating the Eqs.~(\ref{eq:dGamG}) and~(\ref{eq:dGamX}) one obtains the decay width of the Dalitz decay. Note that the widths of the decays considered in this paper have not been measured directly. Instead, the CLEO~\cite{CLEO:2011mla} and BESIII~\cite{BESIII:2021vyq} collaborations measured the ratios of the Dalitz decay with respective to the corresponding radiative decay, namely,
\begin{equation}
	R_{ee}(V) \equiv \frac{\Gamma(V\to Pe^+e^-)}{\Gamma(V\to P\gamma)},
\end{equation}
where the radiative decay width is given by~\cite{Tran:2023hrn}
\begin{equation}
	\label{eq:rad-Gam}
	\Gamma(V\to P\gamma) = \frac{\alpha_{\textrm{em}}}{24}  m_{V}^3\left(1-
	\frac{m_{P}^2}{m_{V}^2}\right)^3g_{VP\gamma}^2(0).
\end{equation} 
Note that the interference of the amplitudes in Eq.~(\ref{eq:M}) is negligible since the X17 boson's total width is very narrow~\cite{Castro:2021gdf}. Therefore, one can simply write
\begin{equation}
	R_{ee}(V) = R^\gamma_{ee}(V)+R^X_{ee}(V).
\end{equation}

The differential decay rate for the case of the X17 mediator can be simplified using the narrow width approximation~\cite{Lee:2025lwv}
\begin{equation}
	\label{eq:narrow}
	\frac{1}{(q^2-m_X^2)+m_X^2\Gamma_X^2} = \frac{\pi}{m_X\Gamma_X}\delta(q^2-m_X^2).
\end{equation}
Additionally, by assuming that the X17 vector boson decays dominantly into $e^+e^-$, one has
\begin{equation}
	\label{eq:GamX}
	\Gamma_X \equiv \frac{e^2\varepsilon_e^2}{12\pi}\left(1+\frac{2m_e^2}{m_X^2}\right)\sqrt{1-\frac{4m_e^2}{m_X^2}}.
\end{equation} 
By substituting Eqs.~(\ref{eq:narrow}) and~(\ref{eq:GamX}) into Eq.~(\ref{eq:dGamX}), the coupling constant $e\varepsilon_e$ cancels out and is therefore not required in the calculation.

\section{Numerical results}
\label{sec:result}
\subsection{Form factors}
The form factors $g_{VP\gamma}(q^2)$ and $g_{VPX}(q^2)$ defined in Eq.~(\ref{eq:T}) are calculated within the CCQM by evaluating the hadronic amplitudes $T_\mu^\gamma$ and $T_\mu^X$. We show below the calculation steps for the case of $g_{VP\gamma}(q^2)$. The calculation of $g_{VPX}(q^2)$ is performed in a similar manner by exchanging $e_{Q,q} \leftrightarrow \varepsilon_{Q,q}$. In the CCQM, the hadronic amplitude $T_\mu^\gamma$ is written as
\begin{eqnarray}
	 \left\langle P(p^\prime)|J_\mu^{\textrm{em}}|V(p,\epsilon_V) \right\rangle &=& (-3i)e g_{V} g_P \epsilon_{V}^\nu(p) (e_Q \mathcal{M}_{\mu\nu}^Q + e_q \mathcal{M}_{\mu\nu}^q),\\
	\mathcal{M}_{\mu\nu}^Q &=& \int \frac{dk}{(2\pi)^4i}\widetilde{\Phi}_{V}\big[-(k-\omega_2p)^2\big]\widetilde{\Phi}_{P}\big[-(k-\omega_2p^\prime)^2\big]\nonumber\\
	&&\times\mathrm{tr}\big[S_q(k)\gamma_\nu S_Q(k-p)\gamma_\mu S_q(k-p^\prime)\gamma^5\big],\\
	\mathcal{M}_{\mu\nu}^q &=& \int \frac{dk}{(2\pi)^4i}\widetilde{\Phi}_{V}\big[-(k+\omega_1p)^2\big]\widetilde{\Phi}_{P}\big[-(k+\omega_1p^\prime)^2\big]\nonumber\\
	&&\times\mathrm{tr}\big[S_q(k+p^\prime)\gamma_\mu S_q(k+p)\gamma_\nu S_Q(k)\gamma^5\big].
\end{eqnarray}
The ratios of quark masses now read $\omega_1=m_{Q}/(m_{Q}+m_{q})$ and $\omega_2=m_{q}/(m_{Q}+m_{q})$ with $Q=b,c$, and $q=u,d,s$. 

Next, we substitute the Gaussian form for the vertex functions in Eq.~(\ref{eq:vertexf}). Note that Feynman diagrams are calculated in the Euclidean region where $p^2=-p^2_E$. The vertex functions fall off in the Euclidean region, and therefore guarantee ultraviolet convergence. The loop integration is performed by using the Fock-Schwinger representation for the quark propagator
\begin{equation}
	S_{q_i} (k) = (m_{q_i} + \not\! k)\int\limits_0^\infty \!\!d\alpha_i\,e^{-\alpha_i (m_{q_i}^2-k^2)}.
	\label{eq:Fock}
\end{equation}

The integrals over the Fock-Schwinger parameters 
$0\le \alpha_i<\infty$ are treated by introducing an additional integration which converts the set of 
these parameters into a simplex as follows
\begin{equation}
	\prod\limits_{i=1}^n\int\limits_0^{\infty} 
	\!\! d\alpha_i f(\alpha_1,\ldots,\alpha_n)
	=\int\limits_0^{\infty} \!\! dtt^{n-1}
	\prod\limits_{i=1}^n \int\!\!d\alpha_i 
	\delta\left(1-\sum\limits_{i=1}^n\alpha_i\right)
	f(t\alpha_1,\ldots,t\alpha_n).
	\label{eq:simplex}  
\end{equation}
At this stage, an infrared cutoff is introduced to avoid any possible thresholds in the Feynman diagram:
\begin{equation}
	\int\limits_0^\infty dt (\ldots) \to \int\limits_0^{1/\lambda^2} dt (\ldots).
	\label{eq:conf}
\end{equation}
The infrared cutoff parameter $\lambda$ effectively guarantees the confinement of quarks within hadrons. 

Finally, the form factor $g_{VP\gamma}(q^2)$ can be written as
\begin{equation}
g_{VP\gamma}(q^2)=e_Q\,I_Q(m^2_{V},m^2_P,q^2)+e_q\,I_q(m^2_{V},m^2_P,q^2),
\end{equation}
where $I_{Q(q)}(m^2_{V},m^2_P,q^2)$ are twofold integrals which are calculated numerically. 
The expression for $I_{Q}(m^2_{V},m^2_P,q^2)$ reads
\begin{eqnarray}
	\label{eq:I}
	I_{Q}(m^2_{V},m^2_P) &=& g_V g_P \frac{N_c}{4\pi^2}\int\limits_0^{1/{\lambda^2}} \!\! 
	\frac{dt\,t^2}{(s+t)^2} \int\limits\!\!d\alpha^3 \delta\big(1-\sum\limits_{i=1}^3\alpha_i \big)\\ \nonumber
	&&\times \left[m_Q\omega_2+m_q\omega_1+\frac{t}{s+t}(m_Q-m_q)(\omega_1-\alpha_2) \right]\exp\left(-tz_0 + \frac{st}{s+t}z_1\right),
	\\[2ex] \nonumber
	z_0 &=&  (1-\alpha_2)m_Q^2+\alpha_2m_q^2-\alpha_1\alpha_2 m_V^2-\alpha_2\alpha_3 m_P^2,
	\nn 
	z_1 &=& m_V^2\left(\alpha_1-\omega_2\frac{s_V}{s}\right)(\omega_1-\alpha_2)+m_P^2(\alpha_2-\omega_1)\left(\alpha_1+\alpha_2-\frac{s_V}{s}-\omega_1\frac{s_P}{s}\right),
	\nn
	s &=& s_V+s_P , \qquad s_{V{(P)}}=1/\Lambda^2_{V{(P)}}. 
	\nonumber
\end{eqnarray}	
The expression for $I_{q}(m^2_{V},m^2_P,q^2)$ can be obtained by simply exchanging $m_Q \leftrightarrow m_q$ and $\omega_1 \leftrightarrow \omega_2$, i.e. $I_{q}(m^2_{V},m^2_P,m_q, m_Q, \omega_1,\omega_2)=I_{Q}(m^2_{V},m^2_P,m_Q, m_q, \omega_2,\omega_1)$.

For convenience in the calculation, we choose a double-pole parametrization to interpolate the calculated form factors as follows
\begin{equation}
	\label{eq:2pole}
	g_{VP\gamma (X)}(q^2)=\frac{g_{VP\gamma (X)}(0)}{1-aq^2+bq^4}.
\end{equation}
This interpolation form well represents the calculated values of all form factors in this study. The parameters of the double-pole interpolation are listed in Table~\ref{tab:FF}.
\begin{table}[htbp]
	\caption{Parameters of the double-pole approximation for $g_{VP\gamma}(q^2)$ and $g_{VPX}(q^2)$ (with $\varepsilon_u = \pm 5.0\times 10^{-4}$ and $\varepsilon_d = \mp 2.9\times 10^{-3}$).}
	\renewcommand{\arraystretch}{0.8}
	\setlength{\tabcolsep}{10pt}
	\begin{center}
		\begin{tabular}{c|ccc|ccc}
			\hline\hline
			\multicolumn{1}{c|}{} &\multicolumn{3}{c|}{$g_{VP\gamma}(q^2)$} 
			&\multicolumn{3}{c}{$g_{VPX}(q^2)$} \\
			\hline
			Transition & $g_{VP\gamma}(0)$ & $a$ & $b$ & $g_{VPX}(0)$ & $a$ & $b$\\
			\hline
			$D^{*+}\to D^+e^+e^-$ & $-0.32$ & 4.02 & 9.33 & $-5.7\times 10^{-3}$ & 2.03 & 1.07\\
			$D^{*0}\to D^0e^+e^-$ & 1.74 & 1.55 & 0.13 & $1.3\times 10^{-3}$ & 1.55 & 0.13\\
			$D^{*+}_s\to D^+_se^+e^-$ & $-0.18$ & 3.90 & 9.63 & $-4.4\times 10^{-3}$ & 1.45 & 0.40\\
			$B^{*+}\to B^+e^+e^-$ & 1.27 & 2.25 & 1.14 & $4.5\times 10^{-4}$ & 4.80 & 14.49\\
			$B^{*0}\to B^0e^+e^-$ & $-0.73$ & 1.96 & 0.43 & $-6.4\times 10^{-3}$ & 1.96 & 0.43\\
			$B^{*0}_s\to B^0_se^+e^-$ & $-0.58$ & 1.29 & $-0.05$ & $-5.1\times 10^{-3}$ & 1.29 & $-0.05$\\
			$J/\psi\to \eta_c e^+e^-$ & 0.64 & 0.14 & 0.003 & $4.8\times 10^{-4}$ & 0.14 & 0.003\\ 
			\hline\hline
		\end{tabular}
	\label{tab:FF}
	\end{center}
\end{table}    

\begin{figure}[htbp]
	\renewcommand{\arraystretch}{0.3}
	\begin{tabular}{ccc}
		\includegraphics[width=0.333\textwidth]{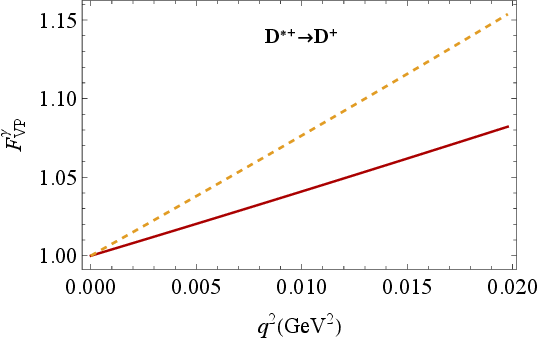}&
		\includegraphics[width=0.333\textwidth]{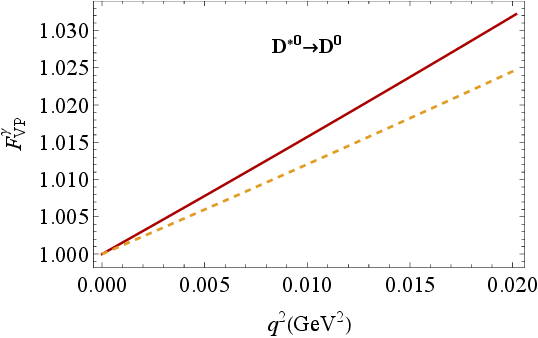}&
		\includegraphics[width=0.333\textwidth]{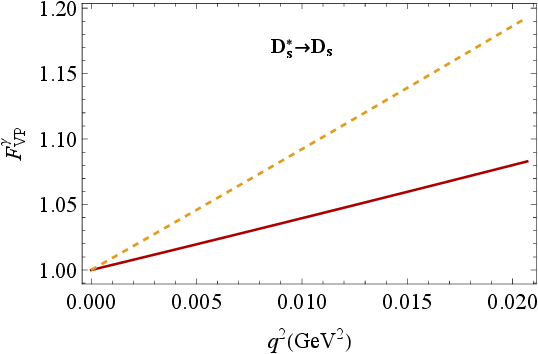}\\
		\includegraphics[width=0.333\textwidth]{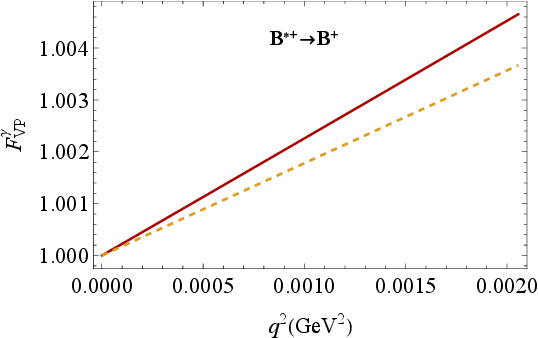}&	
		\includegraphics[width=0.333\textwidth]{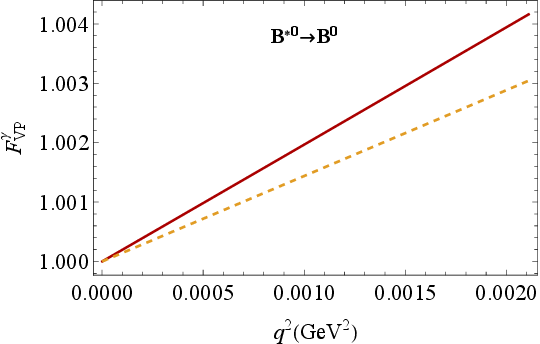}&
		\includegraphics[width=0.333\textwidth]{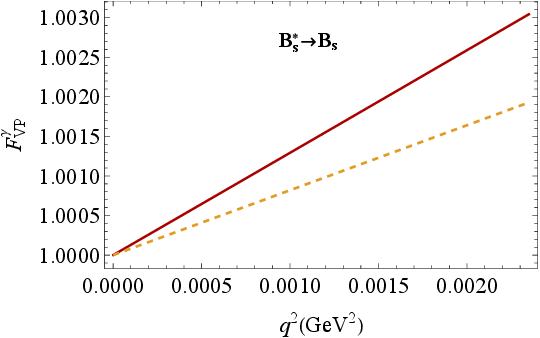}
	\end{tabular}
	\vspace*{-3mm}
	\caption{$F^\gamma_{VP}(q^2)$ in our model (red, solid) and the VMD model (orange, dashed)~\cite{Castro:2021gdf,Lee:2025lwv}.}
	\label{fig:FF-gamma}
\end{figure}

\begin{figure}[htbp]
	\renewcommand{\arraystretch}{0.3}
	\begin{tabular}{ccc}
		\includegraphics[width=0.333\textwidth]{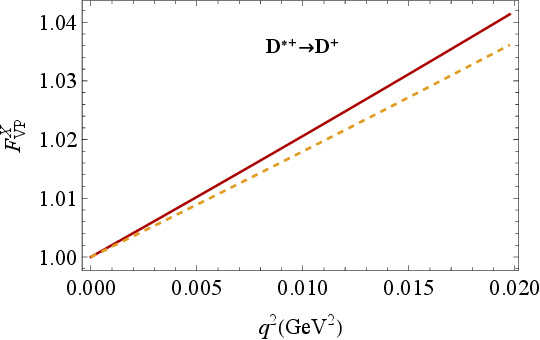}&
		\includegraphics[width=0.333\textwidth]{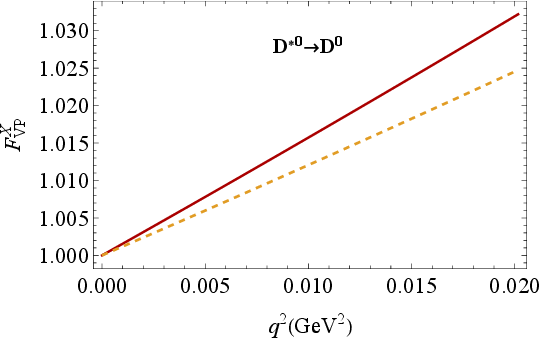}&
		\includegraphics[width=0.333\textwidth]{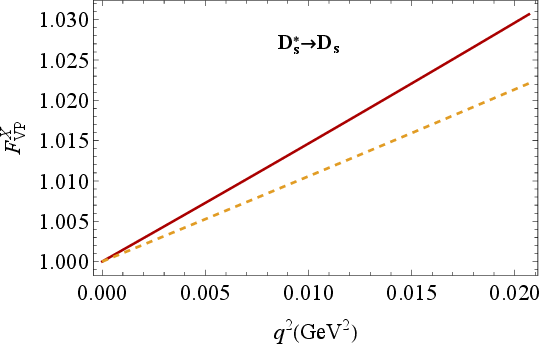}\\
		\includegraphics[width=0.333\textwidth]{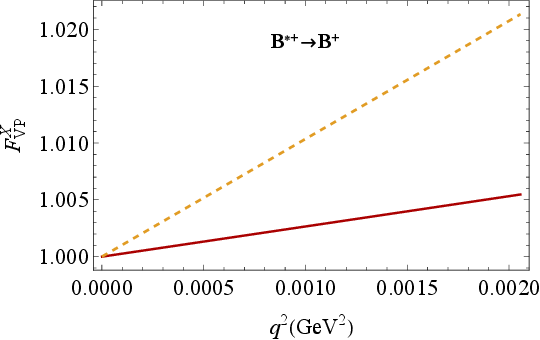}&	
		\includegraphics[width=0.333\textwidth]{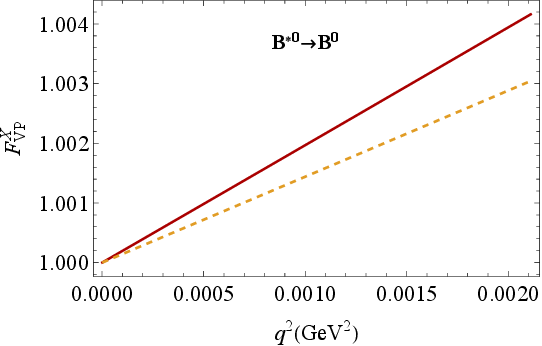}&
		\includegraphics[width=0.333\textwidth]{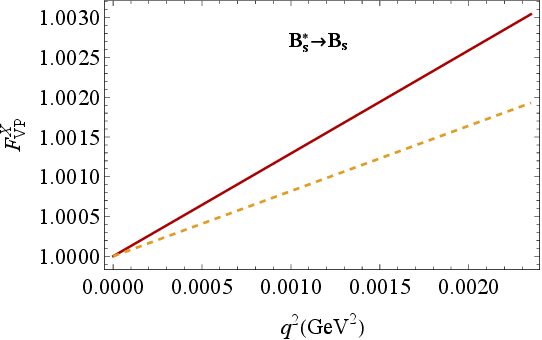}
	\end{tabular}
	\vspace*{-3mm}
	\caption{$F^X_{VP}(q^2)$ in our model (red, solid) and the VMD model (orange, dashed)~\cite{Lee:2025lwv}. The values $\varepsilon_u = \pm 5.0\times 10^{-4}$ and $\varepsilon_d = \mp 2.9\times 10^{-3}$ were used.}
	\label{fig:FF-X}
\end{figure}

\begin{figure}[htbp]
	\renewcommand{\arraystretch}{0.3}
	\begin{tabular}{ccc}
		\includegraphics[width=0.333\textwidth]{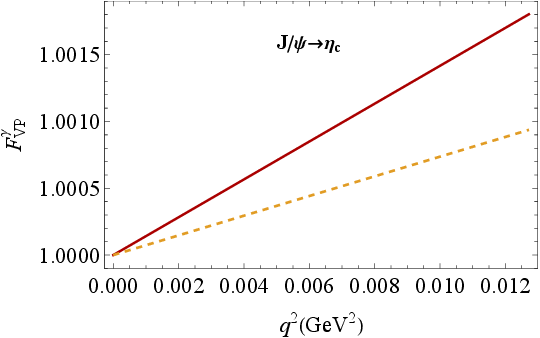}&
		\includegraphics[width=0.333\textwidth]{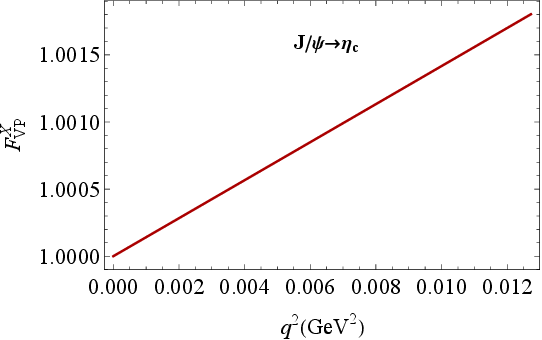}&	
		\includegraphics[width=0.333\textwidth]{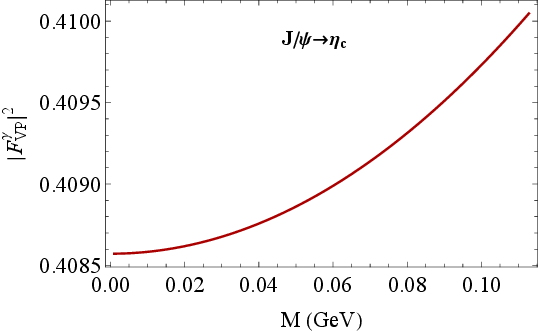}
	\end{tabular}
	\vspace*{-3mm}
	\caption{$J/\psi\to \eta_c e^+ e^-$ form factors in our model (red, solid) and the VMD model (orange, dashed)~\cite{Ban:2020uii}.}
	\label{fig:FF-Jpsi}
\end{figure}

Usually, the form factor $g_{VP\gamma}(q^2)$ is normalized to the corresponding radiative decay constant $g_{VP\gamma}(0)$
\begin{equation}
	F^\gamma_{VP}(q^2) = \frac{g_{VP\gamma}(q^2)}{g_{VP\gamma}(0)}.
	\label{eq:FNormG}
\end{equation}
This normalized form factor appears in Eq.~(\ref{eq:dGamSM}). We can also define a similar normalized form factor for the transition induced by the X17 boson 
\begin{equation}
	F^X_{VP}(q^2) = \frac{g_{VPX}(q^2)}{g_{VPX}(0)}.   
	\label{eq:FNormX}
\end{equation}
In Figs.~\ref{fig:FF-gamma},~\ref{fig:FF-X}, and~\ref{fig:FF-Jpsi}
we plot the normalized form factors calculated in the CCQM and compare them with the form factors obtained using the VMD model.

\begin{figure}[htbp]
	\renewcommand{\arraystretch}{0.3}
	\begin{tabular}{ccc}
		\includegraphics[width=0.333\textwidth]{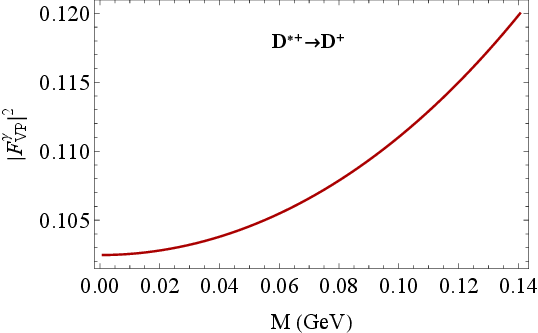}&
		\includegraphics[width=0.333\textwidth]{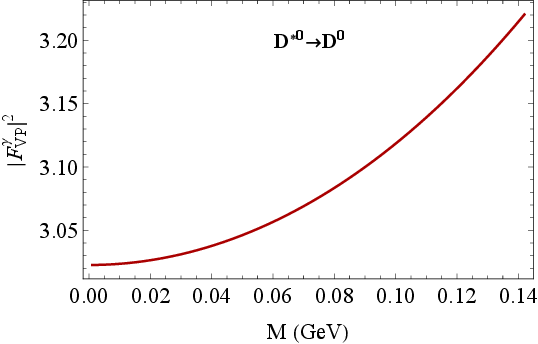}&
		\includegraphics[width=0.333\textwidth]{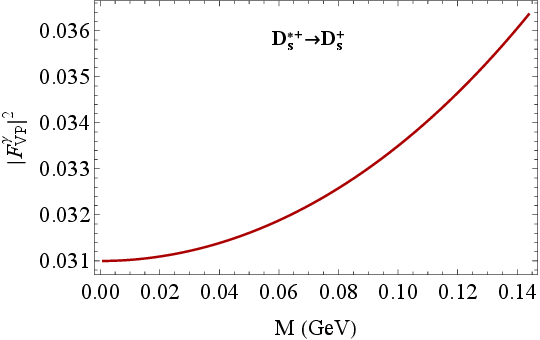}\\
		\includegraphics[width=0.333\textwidth]{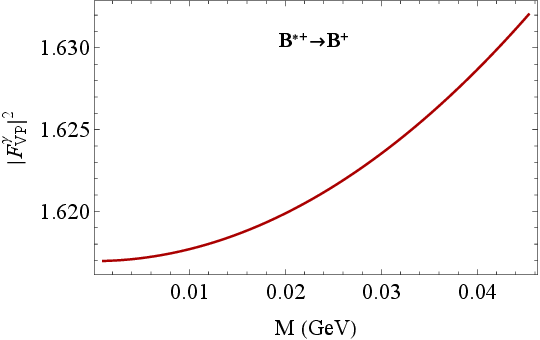}&		
		\includegraphics[width=0.333\textwidth]{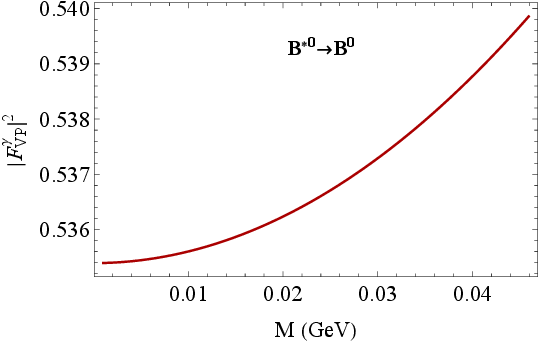}&
		\includegraphics[width=0.333\textwidth]{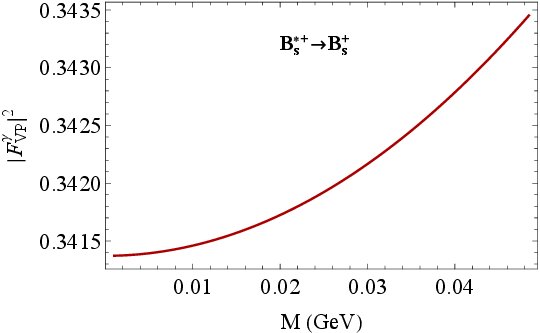}
	\end{tabular}
	\vspace*{-3mm}
	\caption{Normalized form factors squared as functions of the dilepton mass $M = \sqrt{q^2}$.}
	\label{fig:FF-M}
\end{figure}

From an experimental point of view, we can directly measure the squared form factor $|F^\gamma_{VP}(M)|^2$, where $M \equiv \sqrt{q^2}$ is the dilepton mass, by looking at the invariant mass distribution of the lepton pairs produced in the Dalitz decays. We then compare this measured spectrum to the prediction from QED for a point-like interaction~\cite{NA60:2009una}. We therefore also plot $|F^\gamma_{VP}(M)|^2$ in Figs.~\ref{fig:FF-M} and~\ref{fig:FF-Jpsi} for possible comparison with future experiments.  

\subsection{Decay width and the ratio \boldmath{$R_{ee}(V)$}}
Firstly, we present our predictions for the decay widths of the Dalitz decays within the SM. The results are listed in Table~\ref{tab:rate}. The largest decay width belongs to the channel $D^{*0}\to D^0e^+e^-$, which is larger than others by 1--2 orders of magnitude. Note that these rates still have not been measured directly to date. Most often, the ratios $R^\gamma_{ee}(V) =\Gamma(V\to Pe^+e^-)/\Gamma(V\to P\gamma)$ are of the main interest. However, the predictions given here will be important for future experiments, for instance, those at BESIII and the up-coming Super Charm-Tau factory~\cite{Achasov:2024eua}.
\begin{table}[htbp]
\caption{CCQM predictions for the Dalitz decay widths  $\Gamma(V\to Pe^+e^-)$ in the SM (all in eV).}
\renewcommand{\arraystretch}{0.8}
\begin{ruledtabular}
\begin{tabular}{ccccccc}
$D^{*+}\to D^+$ & $D^{*0}\to D^0$ & $D^{*+}_s\to D^+_s$ & $B^{*+}\to B^+$ & $B^{*0}\to B^0$ & $B^{*0}_s\to B^0_s$ & $J/\psi\to \eta_c $\\
\hline
4.03 & 122 & 1.32 &	1.69 & 0.58 & 0.53 & 8.20
\end{tabular}
\end{ruledtabular}
\label{tab:rate}
\end{table}   

Next, we provide the predictions for the ratios $R^\gamma_{ee}(V)$ and $R^X_{ee}(V)$. Note that the ratio $R^X_{ee}(V)$ depends on the coupling $\varepsilon_{Q,q}$ between the X17 boson and the quarks. We will therefore briefly discuss several favored scenarios regarding these couplings. Assuming that the ATOMKI anomaly~\cite{Krasznahorkay:2015iga} is induced by the X17 boson via the process $^8\textrm{Be}^*+\to ^8\textrm{Be}+X(\to e^+e^-)$, the couplings between X17 and the quarks of the first generation are constrained by the condition~\cite{Feng:2016jff,Fornal:2017msy,Feng:2020mbt}
\begin{equation}
|\varepsilon_u+\varepsilon_d|\approx 3.7\times 10^{-3}.
\label{eq:con1}
\end{equation}
The null result from the NA48/2 experimental search for $\pi^0\to X\gamma$ requires the X17 boson to be ``protophobic", i.e. to have a suppressed coupling to the proton~\cite{NA482:2015wmo}
\begin{equation}
|2\varepsilon_u+\varepsilon_d| < 8\times 10^{-4}.
\label{eq:con2}
\end{equation}
By combining the two conditions, one obtains the allowed ranges (i) $\varepsilon_u \in (-4.5,-2.9)$ \& $\varepsilon_d =3.7-\varepsilon_u$ or (ii) $\varepsilon_u \in (2.9,4.5)$ \& $\varepsilon_d =-3.7-\varepsilon_u$. Taking the central values in each range of $\varepsilon_u$, one obtains the favored combination $\varepsilon_u\approx \pm 3.7\times 10^{-3}$ and $\varepsilon_d\approx \mp 7.4\times 10^{-3}$. This combination was used for the calculation in Refs.~\cite{Castro:2021gdf,Ban:2020uii}.
In Table~\ref{tab:R-case1} we present our predictions for the ratios
$R_{ee}^\gamma(V)$, $R_{ee}^X(V)$, and $R_{ee}^{\rm tot}(V)\equiv R_{ee}^\gamma(V)+R_{ee}^X(V)$ using $\varepsilon_u\approx \pm 3.7\times 10^{-3}$ and $\varepsilon_d\approx \mp 7.4\times 10^{-3}$. Note that we assume universal couplings between the X17 boson with up-type and down-type quarks, i.e. $\varepsilon_c = \varepsilon_u$ and $\varepsilon_s =\varepsilon_b = \varepsilon_d$. We also compare with those obtained using the VMD model~\cite{Castro:2021gdf,Ban:2020uii}. Several comments should be addressed:
\begin{table}[htbp]
	\caption{Ratios $R_{ee}^\gamma (V)$ and $R_{ee}^X(V)$ (in units of $10^{-3}$) in the CCQM and VMD model with the couplings $\varepsilon_u =\pm 3.7\times 10^{-3}$ and $\varepsilon_d =\mp 7.4\times 10^{-3}$.}\label{tab:R-case1}
	\renewcommand{\arraystretch}{0.8}
	\begin{ruledtabular}
		\begin{tabular}{cccclc}
			Transition & $R_{ee}^\gamma(V)$ & $R_{ee}^X(V)$ &  $R_{ee}^{\textrm{tot}}(V)$ & Ref. & Experiment \\
			\hline
			$D^{*+}\to D^+e^+e^-$ & $6.47$ & $1.67$ & $8.14 $ & CCQM & {}\\ 
			{} & $6.67$ & $1.05\pm0.07$ & $7.72\pm0.07$ & VMD~\cite{Castro:2021gdf} & {}\\
			$D^{*0}\to D^0e^+e^-$ & $6.45$ & $3.02 \times 10^{-2}$ & $6.48 $ & CCQM & $11.08\pm 0.90$~\cite{BESIII:2021vyq}\\ 
			{} & $6.67$ & $3.02\times 10^{-2}$ & $6.70$ & VMD~\cite{Castro:2021gdf} & {}\\
			$D^{*+}_s\to D^+_se^+e^-$ & $6.50$ & $3.12$ & $9.62 $ & CCQM & $7.2^{+1.8}_{-1.6}$~\cite{CLEO:2011mla}\\ 
			{} & $6.72$ & $3.10\pm0.60$ & $9.82\pm0.60$ & VMD~\cite{Castro:2021gdf} & {}\\
			$B^{*+}\to B^+e^+e^-$ & $4.67$ & $1.77\times 10^{-2}$ & $4.69 $ & CCQM & {}\\ 
			{} & $4.88$ & $(1.91\pm0.03)\times 10^{-2}$ & $4.90$ & VMD~\cite{Castro:2021gdf} & {}\\
			$B^{*0}\to B^0e^+e^-$ & $4.69$ & $3.96\times 10^{-1}$ & $5.09 $ & CCQM & {}\\ 
			{} & $4.88$ & $3.96\times 10^{-1}$ & $5.40$ & VMD~\cite{Castro:2021gdf} & {}\\	
			$B^{*0}_s\to B^0_se^+e^-$ & $5.63$ & $4.05\times 10^{-1}$ & $6.04 $ & CCQM & {}\\ 
			{} & $4.99$ & $4.08\times 10^{-1}$ & $5.40$ & VMD~\cite{Castro:2021gdf} & {}\\
			$J/\psi\to \eta_c e^+e^-$ & $6.08$ & $2.98\times 10^{-2}$ & $6.11 $ & CCQM & {}\\ 
			{} & $6.06$ & $1.32\times 10^{-2}$ & $6.07$ & VMD~\cite{Ban:2020uii} & {}\\			
		\end{tabular}
	\end{ruledtabular}
\end{table}     
\begin{itemize}
	\item 
	Our predictions for $R_{ee}^\gamma(V)$ agree well with those in the VMD model~\cite{Castro:2021gdf,Ban:2020uii} within uncertainty. The central values in the two models differ by less than 5\%, except for the case of $B^{*0}_s\to B^0_se^+e^-$, for which our value is larger by approximately 11\%. Note that in a recent study of charm-meson Dalitz decays~\cite{Tan:2021clg}, the authors also made use of the VMD model and obtained the values $R_{ee}^\gamma(D^{*+})=6.44\times 10^{-3}$,
	$R_{ee}^\gamma(D^{*0})=6.45\times 10^{-3}$, and
	$R_{ee}^\gamma(D^{*+}_s)=6.46\times 10^{-3}$, which are almost the same as our predictions.
	\item 
	Our predictions for $R_{ee}^X(V)$ agree well with those in Ref.~\cite{Castro:2021gdf}, except for the case of $D^{*+}\to D^+e^+e^-$, for which our value is larger by 37\%.
	\item 
	Comparing with Ref.~\cite{Ban:2020uii}, our result for $R_{ee}^\gamma(J/\psi)$ is almost the same as the one in~\cite{Ban:2020uii}. However, our value for $R_{ee}^X(J/\psi)$ is larger by a factor of two.	
	\item 
	The two channels $D^{*+}\to D^+e^+e^-$ and $D^{*+}_{s}\to D^+_{s}e^+e^-$ are most sensitive to the X17 boson contribution. However, the predicted ratio $R_{ee}^{\textrm{tot}}(D^{*+}_s) =9.62\times 10^{-3}$ is slightly larger than the value $7.2^{+1.8}_{-1.6}\times 10^{-3}$ measured by CLEO~\cite{CLEO:2011mla}. 
	\item 
	The X17 boson contribution to the channel $D^{*0}\to D^0e^+e^-$ is smaller than the SM contribution by two orders of magnitude and therefore cannot account for the excess observed recently by BESIII~\cite{BESIII:2021vyq}.
\end{itemize}  
\begin{table}[htbp]
	\caption{Ratios $R_{ee}^\gamma (V)$ and $R_{ee}^X(V)$ (in units of $10^{-3}$) in the CCQM and VMD model with the couplings $\varepsilon_u =\pm 5.0\times 10^{-4}$ and $\varepsilon_d =\mp 2.9\times 10^{-3}$.}\label{tab:R-case2}
	\renewcommand{\arraystretch}{0.8}
	\begin{ruledtabular}
		\begin{tabular}{cccclc}
			Transition & $R_{ee}^\gamma(V)$ & $R_{ee}^X(V)$ &  $R_{ee}^{\textrm{tot}}(V)$ & Ref. & Experiment \\\hline
			$D^{*+}\to D^+e^+e^-$ & $6.47$ & $0.31$ & $6.78 $ & CCQM & {}\\ 
			{} & $6.6$ & $1.2\pm0.4$ & $7.8\pm0.4$ & VMD~\cite{Lee:2025lwv} & {}\\
			$D^{*0}\to D^0e^+e^-$ & $6.45$ & $5.51 \times 10^{-4}$ & $6.45 $ & CCQM & $11.08\pm 0.90$~\cite{BESIII:2021vyq}\\ 
			{} & $6.7$ & $5.6\times 10^{-4}$ & $6.7$ & VMD~\cite{Lee:2025lwv} & {}\\
			$D^{*+}_s\to D^+_se^+e^-$ & $6.50$ & $0.61$ & $7.11 $ & CCQM & $7.2^{+1.8}_{-1.6}$~\cite{CLEO:2011mla}\\ 
			{} & $6.7$ & $4.2\pm 3.1$ & $11\pm 3$ & VMD~\cite{Lee:2025lwv} & {}\\
			$B^{*+}\to B^+e^+e^-$ & $4.67$ & $0.98\times 10^{-4}$ & $4.67 $ & CCQM & {}\\ 
			$B^{*0}\to B^0e^+e^-$ & $4.69$ & $6.08\times 10^{-2}$ & $4.75 $ & CCQM & {}\\ 	
			$B^{*0}_s\to B^0_se^+e^-$ & $5.63$ & $6.22\times 10^{-2}$ & $5.69 $ & CCQM & {}\\ 
			$J/\psi\to \eta_c e^+e^-$ & $6.08$ & $5.43\times 10^{-4}$ & $6.08 $ & CCQM & {}\\ 		
		\end{tabular}
	\end{ruledtabular}
\end{table}     

Recently, Denton and Gehrlein performed a detailed analysis of the X17 vector boson hypothesis~\cite{Denton:2023gat}, taking into account isospin mixing and breaking effects on the  rate for $^8\textrm{Be}^*\to \, ^8\textrm{Be}X$~\cite{Feng:2016ysn}. By using constraints from ATOMKI anomalies in the nuclear decays of excited $^8\textrm{Be}$~\cite{Krasznahorkay:2015iga}, $^4\textrm{He}$~\cite{Krasznahorkay:2021joi}, and $^{12}\textrm{C}$~\cite{Krasznahorkay:2022pxs}, and from the NA48/2 experiment~\cite{NA482:2015wmo}, they obtained the allowed regions $|\varepsilon_u|\approx (0.5 - 0.9)\times 10^{-3}$ and $|\varepsilon_d|\approx (2.5 - 2.9)\times 10^{-3}$ with $\varepsilon_u \varepsilon_d <0$. If isospin effects are ignored, the favored parameters read $|\varepsilon_u| = \pm 5.0\times 10^{-4}$ and $|\varepsilon_d| = \mp 2.9\times 10^{-3}$. If isospin mixing and breaking effects are taken into account, the favored values now become $|\varepsilon_u| = \pm 9.0\times 10^{-4}$ and $|\varepsilon_d| = \mp 2.5\times 10^{-3}$. These two parameter sets were used in Ref.~\cite{Lee:2025lwv} to calculate the X17 contributions to the $D^{*}_{(s)}\to D_{(s)}e^+e^-$ decays within the VMD model. For comparison, we provide our predictions using these parameter sets in Tables~\ref{tab:R-case2} and~\ref{tab:R-case3}. We have the following observations regarding the results in these tables:
\begin{table}[htbp]
	\caption{Ratios $R_{ee}^\gamma (V)$ and $R_{ee}^X(V)$ (in units of $10^{-3}$) in the CCQM and VMD model with the couplings $\varepsilon_u =\pm 9.0\times 10^{-4}$ and $\varepsilon_d =\mp 2.5\times 10^{-3}$.}\label{tab:R-case3}
	\renewcommand{\arraystretch}{0.8}
	\begin{ruledtabular}
		\begin{tabular}{cccclc}
			Transition & $R_{ee}^\gamma(V)$ & $R_{ee}^X(V)$ &  $R_{ee}^{\textrm{tot}}(V)$ & Ref. & Experiment \\\hline
			$D^{*+}\to D^+e^+e^-$ & $6.47$ & $0.21$ & $6.68 $ & CCQM & {}\\ 
			{} & $6.6$ & $0.75\pm0.26$ & $7.4\pm0.3$ & VMD~\cite{Lee:2025lwv} & {}\\
			$D^{*0}\to D^0e^+e^-$ & $6.45$ & $1.8 \times 10^{-3}$ & $6.45 $ & CCQM & $11.08\pm 0.90$~\cite{BESIII:2021vyq}\\ 
			{} & $6.7$ & $1.8\times 10^{-3}$ & $6.7$ & VMD~\cite{Lee:2025lwv} & {}\\
			$D^{*+}_s\to D^+_se^+e^-$ & $6.50$ & $0.40$ & $6.90 $ & CCQM & $7.2^{+1.8}_{-1.6}$~\cite{CLEO:2011mla}\\ 
			{} & $6.7$ & $2.6\pm 1.9$ & $9.3\pm 1.9$ & VMD~\cite{Lee:2025lwv} & {}\\
			$B^{*+}\to B^+e^+e^-$ & $4.67$ & $8.65\times 10^{-4}$ & $4.67 $ & CCQM & {}\\ 
			$B^{*0}\to B^0e^+e^-$ & $4.69$ & $4.52\times 10^{-2}$ & $4.74 $ & CCQM & {}\\ 	
			$B^{*0}_s\to B^0_se^+e^-$ & $5.63$ & $4.62\times 10^{-2}$ & $5.68 $ & CCQM & {}\\ 	
			$J/\psi\to \eta_c e^+e^-$ & $6.08$ & $1.76\times 10^{-3}$ & $6.08 $ & CCQM & {}\\ 		
		\end{tabular}
	\end{ruledtabular}
\end{table}     
\begin{itemize}
	\item 
	Our predictions for $R_{ee}^\gamma(D^*_{(s)})$ fully agree with those calculated in Ref.~\cite{Lee:2025lwv}. The central values in the two studies differ by less than 4\%. Besides, the two studies predict the same value for $R_{ee}^X(D^{*0})$.
	\item 
	Regarding the decays $D^{*+}\to D^+e^+e^-$ and $D^{*+}_{s}\to D^+_{s}e^+e^-$, our predictions for $R_{ee}^X(D^{*+}_{(s)})$ disagree with Ref.~\cite{Lee:2025lwv}. The predicted $R_{ee}^X(D^{*+}_{(s)})$ in Ref.~\cite{Lee:2025lwv} are larger than ours by approximately an order of magnitude. As a result, their total ratios $R_{ee}^{\rm tot}(D^{*+}_{s})$ are slightly larger than the CLEO result~\cite{CLEO:2011mla}. Meanwhile, our values for $R_{ee}^{\textrm{tot}}(D^{*+}_s)$ are consistent with the last. 
	\item 
	The two new parameter sets lead to very small contributions of the X17 boson to the Dalitz decays of $B^*\to B$ and $J/\psi\to \eta_c$. And finally, the X17 boson contribution to the channel $D^{*0}\to D^0e^+e^-$ is now even smaller than in Table~\ref{tab:R-case1} and cannot account for the BESIII excess~\cite{BESIII:2021vyq}.
\end{itemize}

Finally, in Table~\ref{tab:R-case4}, we present our predictions for the decays $D^{*0}\to D^0e^+e^-$, $D^{*+}_s\to D^+_se^+e^-$, and $J/\psi\to \eta_c e^+e^-$ using the best-fit couplings $\varepsilon_u =6.0\times 10^{-2}$, $\varepsilon_c =6.4\times 10^{-3}$, and $\varepsilon_s =-2.0\times 10^{-3}$ recently obtained in Ref.~\cite{Lee:2025lwv}. Note that these values were determined by fitting the VMD predictions for the Dalitz decays of $D_s^*$, $D^{*0}$, $\psi(2S)$, and $\phi$ to available experimental data, assuming that $\varepsilon_u\neq\varepsilon_d\neq\varepsilon_c$ in general. With these coupling parameters, the X17 contributions to $D^{*+}_s\to D^+_se^+e^-$ and $J/\psi\to \eta_c e^+e^-$ are negligibly small. On the other hand, the X17 contributes largely to the decay $D^{*0}\to D^0e^+e^-$ and can explain the BESIII excess. However, the X17 contribution to $D^{*0}\to D^0e^+e^-$ is now approximately equal to the SM one, which seems unlikely. In other words, as already discussed in Ref.~\cite{Lee:2025lwv}, there is a tension between the value of the coupling parameter $\varepsilon_u$ determined from the ATOMKI experiments and the one from the BESIII measurement~\cite{BESIII:2021vyq}.
\begin{table}[htbp]
	\caption{Ratios $R_{ee}^\gamma (V)$ and $R_{ee}^X(V)$ (in units of $10^{-3}$) with the best-fit couplings $\varepsilon_u =6.0\times 10^{-2}$, $\varepsilon_c =6.4\times 10^{-3}$, and $\varepsilon_s =-2.0\times 10^{-3}$~\cite{Lee:2025lwv}.}\label{tab:R-case4}
	\renewcommand{\arraystretch}{0.8}
	\begin{ruledtabular}
		\begin{tabular}{cccccc}
Transition & $R_{ee}^\gamma(V)$ & $R_{ee}^X(V)$ &  $R_{ee}^{\textrm{tot}}(V)$ & Ref. & Experiment \\\hline
$D^{*0}\to D^0e^+e^-$ & $6.45$ & 5.23 & $11.68 $ & CCQM & $11.08\pm 0.90$~\cite{BESIII:2021vyq}\\ 
$D^{*+}_s\to D^+_se^+e^-$ & $6.50$ & $1.87\times 10^{-3}$ & $6.50 $ & CCQM & $7.2^{+1.8}_{-1.6}$~\cite{CLEO:2011mla}\\ 
$J/\psi\to \eta_c e^+e^-$ & $6.08$ & $8.90\times 10^{-5}$ & $6.08 $ & CCQM & {}\\ 		
		\end{tabular}
	\end{ruledtabular}
\end{table}     

\section{Summary}
\label{sec:sum}
Electromagnetic Dalitz decays of the vector mesons $D^{*}_{(s)}$, $B^{*}_{(s)}$, and $J/\psi$ were studied in the framework of the Covariant Confined Quark Model in light of new experimental data from the ATOMKI and BESIII collaborations. Predictions for the form factors and the decay widths (normalized to the corresponding radiative decays) were provided, both in the SM and in the presence of the hypothetical X17 vector boson. We considered four sets of favored coupling constants between the X17 boson and the quarks based on experimental constraints.
Within the SM, we found a full agreement between our predictions and those in other theoretical studies, in which the VMD model was used to obtain the form factors. A detailed comparison of the form factors in the CCQM and in the VMD model was also presented. However, the X17 contributions to the Dalitz decays calculated in the CCQM sometimes disagree with those in the VMD model. The discrepancy can be as large as an order of magnitude. This suggests more independent theoretical calculations are needed.

Finally, among seven decay modes considered in this study, the two decays $D^{*0}\to D^0e^+e^-$ and $D^{*+}_s\to D^+_se^+e^-$ are most sensitive to the X17 contribution. The first has not been observed, but is expected to be measured in the near future by BESIII. Meanwhile, the second was measured by CLEO and the experimental data agrees well with our prediction, especially when the X17 contribution is included. The most interesting decay mode is $D^{*0}\to D^0e^+e^-$, for which theoretical calculations using the CCQM and the VMD model suggest a negligibly small contribution from the X17 (of the order of $10^{-3}\sim 10^{-2}$ compared to the SM contribution). However, the recent BESIII data for this decay showed a $3.5\sigma$ excess over the SM prediction, which cannot be accounted for by the X17 alone. Independent theoretical studies and more experimental data from BESIII for this decay are therefore needed to shed more light on this curious case.

\begin{acknowledgments}
C.T.~Tran and A.T.T.~Nguyen thank HCMC University of Technology and Education for support in their work and scientific collaboration. C.T.~Tran thanks Nestor~Quintero for confirming the value $\varepsilon_e$ used in Ref.~\cite{Castro:2021gdf}. C.T.~Tran thanks Thuc~Uyen~T.L. and her collaborators for fruitful discussion and for providing us with the VMD values given in Table~\ref{tab:R-case3} (Ref.~\cite{Lee:2025lwv}). The authors thank Mauro Raggi and the unknown referee for pointing out relevant references to NA64 and PADME searches for the X17 boson. 
\end{acknowledgments}


\end{document}